\documentclass[aps,twocolumn,showpacs,prl,superscriptaddress]{revtex4}
\usepackage{epsfig,amsmath}

\begin{document}

\title{Counterrotating Dust Disk Around a Schwarzschild Black Hole: \\
New Fully Integrated Explicit Exact Solution}

\author{Guillermo A. Gonz\'{a}lez}
\email[e-mail: ]{guillego@uis.edu.co}
\affiliation{Escuela de F\'{\i}sica, Universidad Industrial de Santander, A.
A. 678, Bucaramanga, Colombia}
\affiliation{Departamento de F\'{\i}sica Te\'orica, Universidad del Pa\'is
Vasco, 48080 Bilbao, Spain}

\author{Antonio C. Guti\'errez-Pi\~{n}eres}
\email[e-mail: ]{gutierrezpac@yahoo.com}
\affiliation{Escuela de F\'{\i}sica, Universidad Industrial de Santander, A. A.
 678, Bucaramanga, Colombia}


\begin{abstract}

The first fully integrated explicit exact solution of the Einstein field
equations  corresponding to the superposition of a counterrotating dust disk
with a central black hole is presented. The obtained solution represents an
infinite annular thin disk (a disk with an inner edge) around the Schwarszchild
black hole. The mass of the disk is finite and the energy-momentum tensor agrees
with all the energy conditions. Furthermore, the total mass of the disk when the
black hole is present is less than the total mass of the disk alone. The
solution can also be interpreted as describing a thin disk made of two
counterrotanting dust fluids that are also in agreement with all the energy
conditions. Additionally, as we will show shortly in a subsequent paper, the
above solution is the first one of an infinite family of solutions.

\end{abstract}

\maketitle

The observational data supporting the existence of black holes at the nucleus of
some galaxies, including the Milky Way, is today so abundant that there is no
doubt about the relevance of the study of binary systems composed by a thin disk
surrounding a central black hole. Accordingly, a lot of work has been developed
in the last years in order to obtain a better understanding of the different
aspects involved in the dynamics of these systems (see \cite{KHS} for a recent
review of the main works). Now, due to the presence of a black hole, the
gravitational fields involved are so strong that the proper theoretical
framework to analytically study this configurations is provided by the general
theory of relativity. Therefore, a strong effort has been dedicated to the
obtention of exact solutions of Einstein equations corresponding to thin
disklike sources with a central black hole. However, until now, any explicit
exact solution corresponding to such kind of systems has been obtained.

Stationary and axially symmetric solutions of the Einstein equations are the
best choice to attempt to describe the gravitational fields of disks around
black holes in an exact analytical manner. At the same time, such spacetimes are
of obvious astrophysical importance, as they describe the exterior of
equilibrium configurations of bodies in rotation. So, through the years, several
examples of solutions corresponding to black holes or to thin disklike sources
has been obtained by many different techniques. However, due to the nonlinear
character of the Einstein equations, solutions corresponding to the
superposition of black holes and thin disks are not so easy to obtain and so,
until now, exact stationary solutions have not been obtained.

On the other hand, if we only consider static configurations, the line element
is characterized only by two metric functions. Furthermore, in the vacuum case,
the Einstein equations system implies that one of the metric functions satisfies
the Laplace equation whereas that the other one can be obtained by quadratures.
Therefore, as a consequence of the linearity of the Laplace equation, solutions
corresponding to the superposition of thin disks and black holes can be, in
principle, easily obtained. However, only very few solutions has been obtained
and neither of them has been fully explicitly integrated \cite{LL1, LL2, LL3,
SZ1, SEM1, SZ2, SEM2, SEM3}.

In this letter we present what seems to be the first fully integrated explicit
exact solution for the superposition of a thin disk and a black hole. We begin
by considering the Weyl metric for a static axially symmetric spacetime, written
as \cite{KSHM}
\begin{equation}
ds^2 = - e^{2\Phi}dt^2 + e^{-2\Phi}[r^2d\varphi^2 + e^{2\Lambda}(dr^2 + dz^2)],
\label{eq:met}
\end{equation}
with $\Phi$ and $\Lambda$ only depending on $r$ and $z$. The Einstein vacuum
equations leads to the Laplace equation for $\Phi$,
\begin{equation}
\Phi_{,rr} + \frac{1}{r} \Phi_{,r} + \Phi_{,zz} = 0, \label{eq:weyl1} 
\end{equation}
and, given $\Phi$, $\Lambda$ is obtained by solving the quadrature
\begin{equation}
\Lambda[\Phi]  = \int r [ ( \Phi_{,r}^2 - \Phi_{,z}^2 ) dr + 2 \Phi_{,z}
\Phi_{,z} dz ], \label{eq:weyl2}
\end{equation}
whose integrability is granted by equation (\ref{eq:weyl1}).

We consider a solution of (\ref{eq:weyl1}) of the form
\begin{equation}
\Phi = \psi + \phi, \label{eq:phi}
\end{equation}
where $\psi$ corresponds to a black hole solution whereas that $\phi$
corresponds to a thin disk solution. Now, by using (\ref{eq:phi}) in
(\ref{eq:weyl2}), we obtain
\begin{equation}
\Lambda[\Phi] = \Lambda[\psi] + \Lambda[\phi] + 2 \Lambda[\psi,\phi],
\label{eq:lam}
\end{equation}
with
\begin{eqnarray}
\Lambda[\psi,\phi]  &=& \int r [ ( \psi_{,r}  \phi_{,r} - \psi_{,z} \phi_{,z} )
dr \nonumber \\
&&+ ( \psi_{,r} \phi_{,z} + \psi_{,z} \phi_{,z} ) dz ], \label{eq:lamix}
\end{eqnarray}
a term due to the nonlinear character of (\ref{eq:weyl2}). For the black hole we
take $\psi$ and $\Lambda[\psi]$ as given by the Schwarszchild solution written
as
\begin{eqnarray}
\psi &=& \frac{1}{2} \ln \left[ \frac{\zeta - 1}{\zeta + 1} \right], \\
\Lambda[\psi] &=& \frac{1}{2} \ln \left[ \frac{\zeta ^2 - 1}{\zeta ^2 - \eta^2}
\right],
\end{eqnarray}
where the prolate spheroidal coordinates are defined by means of 
\begin{equation}
r^2 = m^2 (\zeta^2 - 1)(1 - \eta^2), \qquad z = m \zeta  \eta
\end{equation}
and $1 \leq \zeta  < \infty$, $- 1 \leq \eta \leq 1$.

Now, in order to obtain the thin disk solution, we introduce the oblate
spheroidal coordinates by means of
\begin{equation}
r^2 = a^2 (x^2 + 1)(1 - y^2), \qquad z = a x y,
\end{equation}
with the ranges taken as $- \infty < x < \infty$, $0 \leq y \leq 1$. The disk is
obtained by taking $y = 0$ and so is located at $z = 0$, $r \geq a$. On crossing
the disk, $x$ changes sign but does not change in absolute value, so that an
even function of $x$ is a continuous function everywhere but has a discontinuous
$x$ derivative at the disk. By solving the Laplace equation (\ref{eq:weyl1}),
with the proper boundary conditions corresponding to a thin disk with an inner
edge, we obtain for $\phi$ the simple expression
\begin{equation}
\phi = \frac{\alpha y}{a (x^2 + y^2)}.
\end{equation}
Then, after a simple integration of (\ref{eq:weyl2}), we obtain for
$\Lambda[\phi]$ the expression
\begin{equation}
\Lambda[\phi] = - \frac{\alpha^2 (1 - y^2) A(x,y)}{4 a^2 (x^2 + y^2)^4},
\end{equation}
where
$$
A(x,y) = x^4 (9 y^2 - 1) + 2 x^2 y^2 (y^2 + 3) +
y^4 (y^2 - 1),
$$
with $\alpha$ an arbitrary constant and $a$ the inner radius of the disk.
\begin{figure}[t]
\begin{center}\epsfig{width=2.75in,file=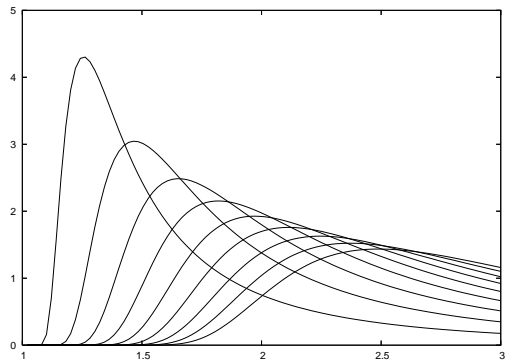}\end{center}
\caption{Energy density ${\tilde \epsilon} = a \epsilon$ as a function of
${\tilde r} = r/a$ for ${\tilde \alpha} = \alpha/a = 1, ... , 9$. The first
curve on left corresponds to ${\tilde \alpha} = 1$, whereas that the last curve
on right corresponds to ${\tilde \alpha} = 9$.}\label{dendis}
\end{figure}

The surface energy-momentum tensor of the disk can be written as \cite{GE}
\begin{equation}
S_{ab} \ = \ \epsilon V_a V_b , \label{eq:emt0}
\end{equation}
where $V^a = e^{- \Phi} \delta_{_0}^a$. The surface energy density is given by
\begin{equation}
\epsilon = \frac{4 \alpha}{a^2 x^3} \exp \left\{- \frac{\alpha^2}{4 a^2
x^4} \right\} ,
\end{equation}
for $x \geq 0$, and will be allways positive if we take $\alpha > 0$. We then
have a dust disk in agreement with all the energy conditions. The total mass of
the disk can be easily computed and we obtain
\begin{equation}
\frac{M}{2 \pi} = \int_a^\infty \epsilon (r) r dr = \sqrt{2 a \alpha} \ \Gamma
(1/4),
\end{equation}
so that the disk is of infinite extension but with finite mass. In Fig.
\ref{dendis} we plot the energy density for some values of $\alpha$.

For the combined black hole and disk system, the integration of the mixed term
$\Lambda[\phi,\psi]$ in (\ref{eq:lam}) can be done with aid of some very useful
computational techniques introduced in \cite{LO}. So, we obtain
\begin{equation}
\Lambda[\phi,\psi] = \frac{\alpha}{2} \left[ \frac{1 - y}{x^2 + y^2}
\right] \left( \Lambda_1 - \Lambda_2 \right) ,
\end{equation}
where
\begin{eqnarray*}
\Lambda_1 &=& \frac{a y (1 - y) (1 + x^2) - m x (1 + y) (\zeta  - 1)(1 -
\eta)}{[a x + m (1 - \zeta  - \eta) ]^2 + a^2 (1 - y)^2 }, \\
	& &	\\
\Lambda_2 &=& \frac{a y (1 - y) (1 + x^2) - m x (1 + y) (\zeta  + 1)(1 -
\eta)}{[a x - m (1 + \zeta  - \eta) ]^2 + a^2 (1 - y)^2 },
\end{eqnarray*}
that vanishes at the axis, when $y = 1$, and at the disk, when $y = \eta = 0$.
 More
details about the obtention of the solution will be presented in a subsequent
paper.

The presence of the black hole modifies the energy-momentum tensor of the disk,
in such a way that arises a nonzero pressure at the azimuthal direction.
Accordingly, the modified energy-momentum tensor of the disk can be written as
\cite{GE}
\begin{equation}
{\cal S}_{ab} = \varepsilon V_a V_b + p X_a X_b, \label{eq:emt}
\end{equation}
where $X^a = e^{\Phi} \delta_{_1}^a$. The energy density, the azimuthal pressure
and the ``effective Newtonian density'' are given, respectively,  by
\begin{eqnarray}
\varepsilon &=& \left[ \frac{\zeta  - 1}{\zeta  + 1} \right] \epsilon, \\
	&&	\nonumber \\
p &=& \left[ \frac{1}{\zeta  + 1} \right] \epsilon, \\
	&&	\nonumber \\
\sigma &=& \left[ \frac{\zeta }{\zeta  + 1} \right] \epsilon.
\end{eqnarray}
where $\sigma = \varepsilon + p$. Now, as $\epsilon \geq 0$ and $\zeta \geq 1$,
$\varepsilon$ and $\sigma$ will be positives everywhere. Accordingly, the
energy-momentum tensor will be in fully agreement with the weak and strong
energy conditions. On the other hand, in order to fulfill the dominant energy
condition, we require that $p \leq \varepsilon$, which implies that $\zeta \geq
2$ and $a \geq \sqrt{3} m$. Furthermore, we have that $\sigma < \epsilon$, and
thus the total mass of the disk when the black hole is present is less than the
total mass of the disk alone.

The energy-momentum tensor can also be interpreted as the superposition of two
counterrotating fluids. In order to do this, we cast ${\cal S}^{ab}$ as
\cite{GE}
\begin{equation}
{\cal S}^{ab} = \varepsilon_+ U_+^a U_+^b + \varepsilon_- U_-^a U_-^b,
\label{eq:crm}
\end{equation}
where
\begin{equation}
\varepsilon_+ = \varepsilon_- = \left[ \frac{\zeta  - 2}{\zeta  + 1} \right]
\frac{\epsilon}{2},
\end{equation}
are the energy densities of the two counterrotating fluids. The counterrotating
velocity vectors are given by \cite{GE}
\begin{equation}
U_\pm^a = \frac{V^a \pm U X^a}{\sqrt{1 - U^2}},
\end{equation}
where
\begin{equation}
U^2 = \frac{p}{\varepsilon} = \frac{1}{\zeta  - 1} \leq 1,
\end{equation}
is the counterrotating tangential velocity. So, we have two counterrotating dust
fluids with equal energy densities.  Now, as $\varepsilon_\pm \geq 0$, the two
counterrotating dust disks are in fully agreement with all the energy
conditions.

As we can see, the above solution presents some very interesting properties.
First, the associated material source presents a very reasonable behavior and
its energy momentum tensor is in agreement with all the energy conditions.
Furthermore, its relative simplicity when expressed in terms of the spheroidal
coordinates, prolates and oblates, makes it very easy to study different
dynamical aspects, like the motion of particles inside and outside the disk, the
stability of the orbits and the possible existence of singularities.
Additionally, as we will show shortly in a subsequent paper, the above solution
is the first one of an infinite family of solutions, all of them having many
remarkable properties in common with the solution here presented.

{\it Acknowledgments}. A. C. G-P. wants to thank the financial support from
COLCIENCIAS, Colombia.


\begin{thebibliography}{9999}

\bibitem{KHS} V. Karas, J-M. Hure and O. Semer\'ak. Class. Quantum Grav. {\bf
21}, R1-R51 (2004)

\bibitem{LL1} J. P. S. Lemos and P. S. Letelier, Class.  Quantum Grav. {\bf
10}, L75 (1993).

\bibitem{LL2} J. P. S. Lemos and P. S. Letelier, Phys. Rev D  {\bf 49}, 5135
(1994).

\bibitem{LL3} J. P. S. Lemos and P. S. Letelier, Int. J.  Mod.  Phys. D {\bf
5}, 53 (1996).

\bibitem{SZ1} O. Semer\'ak and M. \u{Z}\'ac\u{e}k, Class. Quantum Grav. {\bf
17}, 1613 (2000)

\bibitem{SEM1} O. Semer\'ak, Class. Quantum Grav. {\bf 19}, 3829 (2002)

\bibitem{SZ2} M. \u{Z}\'ac\u{e}k and O. Semer\'ak, Czech. J. Phys. {\bf 52}, 19
(2002)

\bibitem{SEM2} O. Semer\'ak, Class. Quantum Grav. {\bf 20}, 1613 (2003)

\bibitem{SEM3} O. Semer\'ak, Class. Quantum Grav. {\bf 21}, 2203 (2004)

\bibitem{KSHM} H. Stephani, D. Kramer, M. McCallum, C. Hoenselaers and E. Herlt,
and  {\it Exact Solutions to Einsteins's  Field Equations} (Second Edition,
Cambridge University Press, Cambridge, England, 2003).

\bibitem{GE} G. A. Gonz\'alez and O. A. Espitia,  Phys. Rev. D  {\bf 68},
104028 (2003).

\bibitem{LO} P.S. Letelier and S. R. Oliveira, J. Math. Phys. {\bf 28}, 165
(1987).

\end{thebibliography}
\end{document}